\documentclass[manuscript]{aastex}

\begin{document}
\pagenumbering{arabic}

\title{LENTICULAR GALAXIES AND THEIR ENVIRONMENTS}

\author{Sidney van den Bergh}
\affil{Dominion Astrophysical Observatory, Herzberg Institute of Astrophysics, National Research Council of Canada, 5071 West Saanich Road, Victoria, BC, V9E 2E7, Canada}
\email{sidney.vandenbergh@nrc-cnrc.gc.ca}

\begin{abstract}   

    It is widely believed that lenticular (S0) galaxies were initially spirals from which the gas has been removed by interactions with hot cluster gas, or by ram-pressure stripping of cool gas from spirals that are orbiting within rich clusters of galaxies. However, problems with this interpretation are that:
(1) Some lenticulars, such as NGC 3115, are isolated field galaxies rather than cluster members. (2) The distribution of flattening values of S0 galaxies in clusters, in groups and in the field are statistically indistinguishable. This is surprising because one might have expected most of the progenitors of field S0 galaxies to have been flattened late-type galaxies, whereas lenticulars in clusters are thought to have mostly been derived from bulge-dominated early-type galaxies. (3) It should be hardest for ram-pressure to strip massive luminous galaxies with deep potential wells.
However, no statistically significant differences are seen between the luminosity distributions of early-type Shapley-Ames galaxies in clusters, groups and in the field. (4) Finally both ram-pressure stripping and evaporation by hot intra-cluster gas would be most efficient in rich clusters. However, the small number of available data in the Shapley-Ames sample appears to show no statistically significant differences between the relative frequencies of dust-poor $S0_{1}$ and dust-rich $S0_{3}$ galaxies in clusters, groups and in the field. It is tentatively concluded that ram-pressure stripping, and heating by intra-cluster gas, may not be the only evolutionary channels that lead to the formation of lenticular galaxies. It is speculated that gas starvation, or gas ejection by active nuclei, may have play a major role in the formation of a significant fraction of all S0 galaxies.

\end{abstract}

\keywords{galaxies: clusters: general}

\section{INTRODUCTION}

    Hubble (1936,p.45) introduced the S0 classification type as a ``more or less hypothetical'' transitional stage between elliptical and spiral galaxies. In his famous tuning fork diagram they form the missing link between the E, S and SB branches of his classification scheme. Later Spitzer \& Baade
(1951) suggested that S0 (lenticular) galaxies had been formed by the removal of gas during collisions between galaxies in rich clusters of galaxies. However, Gunn \& Gott (1972) pointed out that more modern cluster parameters ``make it somewhat unlikely''
that collisions between galaxies in clusters are sufficiently frequent to account for the observed number of S0 galaxies in clusters. They therefore proposed that the lenticular galaxies in clusters could have been formed by the ram-pressure stripping of interstellar gas and dust from spirals, as they plowed through hot intra-cluster gas. Alternatively (Moore et al. 1996) multiple high-speed interactions in clusters (``galaxy harassment'') might drive the morphological evolution of galaxies in clusters.
Such scenarios for the formation of lenticular galaxies have been widely accepted in recent decades. However, as Baade (1963, p.79) pointed out long ago, there is one big problem: ``How are we to explain the S0 galaxies that are observed in the general field?''
Perhaps the most famous example of such a lenticular field galaxy is NGC 3115, which apart from its dE,N companion NGC 3115 DW1 (Puzia et al. 2000), is an isolated field galaxy. The presence of a 2 x $10^{9}$ $M_{\odot}$ black hole at the center of this object (Kormendy et al. 1996) encourages the speculation that the disk gas might have been removed from this object during an early  quasar-like event. This suggests that active nuclei might have played an important role in the transformation of spiral galaxies into lenticulars.

   The morphological evolution of the earliest type (S0 and E) galaxies is an important, and sometimes controversial, subject.
For example Holden et al. (2009) find no evolution in the mean ellipticities of early-type galaxies between z $\sim$ 1 and z $\sim$ 0. Since S0 galaxies are, on average, much more flattened than ellipticals this would seem to suggest that the E to S0 ratio has not changed significantly over the last $\sim$ 8 Gyr.
Furthrmore van der Wel et al. (2007) find that the fraction of E + S0 galaxies does not appear to have changed significantly over the redshift range 0.03 $<$ z $<$ 0.8. On the other hand Poggianti et al. (2009) find that, over the range z = 1.2 to z = 0.5, the spiral and S0 fractions evolved significantly in clusters with $\sigma$ $<$ 800 km $s^{-1}$ (but not in clusters with a larger velocity dispersion). Holden et al. (2007) have suggested that the evolutionary transformation of S0 galaxies might be strongly dependent on their mass. Earlier Dressler et al. (1997) had concluded that S0s are generated in large numbers only after cluster virialization. A critical problem in all investigations of this type is that it is often {\it very} difficult to unambiguously distinguish E and S0 galaxies. The distinction between these two types of galaxies is most easily seen in objects that are viewed almost edge-on. As a result there is a tendency by some morphologists to mis-classify lenticular galaxies viewed pole-on as ellipticals. Furthermore the effects of decreasing resolution pose a serious challenge to those who attempt to classify (typically rather compact) early-type galaxies at large redshifts. All classification problems of this type are minimized in nearby galaxies that have been observed with large telescopes. The golden standard for such investigations is provided by the relatively nearby galaxies in A Revised Shapley-Ames Catalog of Bright Galaxies (Sandage \& Tammann 1981). These objects have been observed with the large reflectors on Mt. Wilson, Palomar and Las Campanas, and they have been classified in a homogeneous fashion by two highly experienced galaxy morphologists. In a previous paper van den Bergh (2002) those Shapley-Ames galaxies, that are located north of $\delta$ = $-27^{o}$, were assigned to either (1) clusters, (2) groups, or (3) field environments. It is the purpose of the present investigation to look into possible relations between galaxy morphology and environment in the Shapley-Ames sample. It is hoped that this may throw some light on possible links between the morphological evolution of early-type galaxies and their environment.

\section{LUMINOSITY DISTRIBUTION OF E, S0 AND Sa GALAXIES}

   Table 1, which is based on data compiled in Sandage \& Tammann (1981) and van den Bergh (2002), shows that E0 and Sa galaxies are, on average, about twice as luminous as S0 galaxies. This difference is found to hold in all environments. A Kolmogorov-Smirnov test performed on the data listed in this table shows that the probability of E and S0 galaxies having been drawn from the same parent luminosity distribution is 0.2\%, 1\% and 7\% respectively, in clusters, groups and the field. In other words, elliptical galaxies are more luminous than lenticulars in all environments. [The S0 galaxies in the particularly rich Virgo cluster are also fainter than those of types E and Sa, but (because of their small number) not at a respectable level of statistical significance.] Furthermore the data in Table 1 also show that the probabilities of S0 and Sa galaxies in clusters, groups and in the field, having been drawn from the same parent luminosity distribution are only 8\%, not meaningful and 3\%, respectively. It is noted in passing that the luminosity distributions of S0 and SB0 galaxies in Sandage \& Tammann (1981) do not differ significantly. By the same token no significant difference is found between the luminosity distributions of Sa and SBa galaxies. Recently Nair (2009, p.76) has shown that, among 13575 Sloan Digital Sky Survey galaxies, objects of type S0 are also fainter than are those of morphological types E and Sa. The conclusion that the luminosities of S0 galaxies are not intermediate between those of types E and Sa therefore isn't just a quirk of the relatively small Shapley-Ames sample. Burstein et al. (2005) have also found that elliptical galaxies are more luminous than lenticulars from K-band photometry of E and S0 galaxies in the Third Reference Catalogue of Bright Galaxies by de Vaucouleurs et al. (1991). This observed luminosity difference between E and S0 galaxies cannot be blamed on evolutionary effects. This is so because the intrinsic colors of E and S0 galaxies are observationally indistinguishable. Sandage \& Visvanathan (1978) find that the fully corrected mean colors of field + cluster galaxies are $<u - r> = 2.83 \pm  0.01$ and $<u - r> = 2.85 \pm  0.01$ respectively , for S0 and E galaxies.
    
In van den Bergh (2009) it was speculated that the observation that S0 galaxies are, on average, 1.0 mag and 0.8 mag fainter, respectively, than E and Sa galaxies, indicates that typical lenticular galaxies have lost about half of their initial luminous mass. Alternatively one might, of course, argue that faint low-mass spirals are more likely to have been stripped of gas by ram pressure than are luminous high-mass spirals. In other words both location in the cluster environment, or low parent galaxy mass, might have favored the transformation of spirals into lenticulars. However, if ram-pressure stripping (Gunn \& Gott 1972) had transformed cluster Sa galaxies into lenticulars one would have expected the fraction of all early-type galaxies that are assigned to the S0 class would be higher in clusters than it is in the field.  Unexpectedly, the data in Table 1 show that the fraction of all E + S0 + Sa galaxies that are lenticulars is 40/121 (33\%) for clusters, 17/45 (38\%) in groups, and 24/70 (34\%)in the field. In other words the fraction of all early-type galaxies that are of type S0 appears to be more-or-less independent of environment. However, the reader is cautioned that this conclusion is based on a rather small data sample and should be re-investigated when larger databases becomes available. In particular, it would be very important to check if the fraction of all early-type (E+S0+Sa) galaxies that is of type S0 is larger in very rich clusters than it is in other environments. In summary, it is found that (on average) S0 galaxies are (in all environments) only half as luminous as objects of types E and Sa. In this sense S0 galaxies are therefore not, as Hubble (1936) suggested, intermediate between types E and Sa. This conclusion is independently confirmed by the work of Nair (2009, p.74) who shows that, in a sample of 13534 Sloan Digital Sky Survey images, the Petrosian (1976) radii of S0 galaxies are significantly smaller than are those of both E and Sa galaxies.

\section{FLATTENING OF S0 GALAXIES}

    Sandage \& Tammann (1981) list eye estimates of the axial ratios of both the E and S0 galaxies in their catalog. These estimates were used to derive  values of e = 10 (a-b)/a. Their data on the flattening of S0 galaxies in various environments are summarized in Table 2 As expected the S0 galaxies are, on average, much more flattened than are the ellipticals. A Kolmogorov-Smirnov test shows that this conclusion holds true in both clusters (P = 99.9\%), in groups (P = 99.8\%) and in the fieled (P = 97\%). If elliptical galaxies form from merged spirals (Toomre 1977) then one might perhaps have expected their angular momenta (and hence their flattenings) to depend on environment. However, a K-S test of the ellipticities of galaxies in the present sample shows no significant differences between the ellipticity distributions of E galaxies in the cluster, group and field environments. A caveat is, of course, that the present sample of
98 ellipticals is small. Among the 75 lenticular galaxies listed in Table 2 there is also no evidence for a significant difference in the distribution of S0 flattenings between the cluster, group and field environments. In summary; the distribution of flattening values in both E and S0 galaxies appears to be broadly independent of environment. It is also noted in passing that the flattening of lenticular galaxies does not appear to depend on luminosity for objects of types S0(0) - S0(8). However, the sparse data in the Shapley-Ames catalog do hint at the possibility that the most flattened lenticulars, i. e. objects of types S0(9) and S0(10), may be of below-average luminosity.
    
If lenticular galaxies form by sweeping gas from spirals, as suggested by Spitzer \& Baade (1951) and Gunn \& Gott (1972), then one would have expected the majority of intrinsically flattened S0 galaxies to have formed from swept Sc spirals, whereas less intrinsically flattened lenticulars would mostly be expected to have evolved by removal of gas from Sa galaxies. This is so because early-type spirals tend to predominate in clusters, whereas late-type galaxies dominate the field . One might therefore have predicted that the S0 galaxies in the field would, on average, be flatter than those in clusters. To check this the observed (Sandage \& Tammann 1981) flattenings of lenticular galaxies are tabulated for various environments in Table 2. In compiling these data intermediate types such as S0/SB0, E/S0 and S0/Sa were omitted. Surprisingly a Kolmogorov-Smirnov test shows no significant difference between the distribution of flattening values for lenticular galaxies in galaxies that Sandage \& Tammann assign to the Virgo cluster, to S0 galaxies in all clusters, to those in groups and for those assigned to the field. It is particularly striking (and unexpected) that lenticular galaxies in the dense Virgo cluster appear to have a flattening distribution similar to that of lenticulars in the field.
It should, however, be emphasized that this conclusion is based on only a small number of galaxies. Clearly it would be important to try to extend this type of analysis to a larger database, such as that provided by the morphological classifications of thousands of nearby SDSS galaxies by Nair (2009).
    
Theoretically (e.g. McCarthy et al. 2008) one would expect the stripping efficiency to be proportional to the ram pressure, and inversely proportional to the mass of the galaxy being stripped.
Unfortunately the present data are not numerous enough to use a two-dimensional K-S test (Peacock 1983, Fasano \& Franceschini 1987) to check if there is indeed a significant dependence of environment on the location of S0 galaxies in the luminosity versus apparent flattening plane. In summary presently available data provide no evidence for a dependence of the flattening, or luminosity, of galaxies of types S0(0) to S0(8) on environment.

\section{SUBTYPES OF LENTICULAR GALAXIES}

    Following Sandage (1961) and Sandage \& Tammann (1981) lenticular galaxies may be divided into subtypes $S0_{1}, S0_{2}$ and $S0_{3}$, where objects of type $S0_{1}$ are dust-free, and those of type $S0_{3}$ may exhibit a considerable amount of internal dust extinction. Although small in number, the data listed in Table 3 appear to show no statistically significant differences in the distributions of S0 subtypes as a function of environment. Such differences would have been expected if ram-pressure stripping of gas and dust had been more effective in clusters than in the field. It is also noted in passing that there appears to be no statistically significant difference between the relative frequency distribution of S0 subtypes in the giant Virgo cluster and among Shapley-Ames galaxies that are located in lesser clusters.

\section{ FREQUENCY OF S0 GALAXIES}

    Table 4 lists the frequency distribution of various morphological types in the Shapley-Ames catalog as a function of environment. The data were drawn from the environmental designations listed in van den Bergh (2002), supplemented by the assignment of galaxies to the Virgo cluster listed in Sandage \& Tammann (1981). These data strengthen and confirm the trends seen by many previous authors who found that lenticular galaxies are most common in clusters. Specifically it is found that the fraction of all galaxies that are of types S0 or SB0 is 8\% in the field, 13\% in groups, 15\% in all clusters and 28\% in the dense Virgo cluster. The two main conclusions to be drawn from these data are that : (1) The fraction of all galaxies that are lenticular increases with environmental density and (2) the abundance of S0 galaxies among field galaxies is non-negligible. These results suggest that spiral galaxies might have evolved into lenticulars via multiple channels i. e. via (1) ram-pressure stripping in clusters, (2) interactions with hot intra-cluster gas, (3) harassment in dense environments or (4) internal processes, such as winds and star formation induced by active galactic nuclei.

\section{ DISCUSSION}

    Present ideas on the origin of S0 galaxies are mainly based on the early investigations by Spitzer \& Baade (1951), Gunn \& Gott (1972) and Moore et al. (1996). Furthermore (Bekki et al.
2002) hydrodynamical interactions between spiral galaxies and hot intra-cluster gas might result in gas starvation for cluster spiral galaxies. However, these papers do not explain two important
observations: (1) Significant numbers of S0 galaxies exist in the field, where ram-pressure stripping, contact with hot intra-cluster gas  and harassment should be unimportant, and (2) The luminosity distribution of S0 galaxies (see Table 1) shows no significant differences between the cluster, group and field environments. Furthermore,(3) ram-pressure stripping should be less effective in massive luminous S0 galaxies than in faint less massive ones. However, the data in Table 1 appear to show no statistically significant dependence of the luminosity distribution of lenticular galaxies on environment. Finally (4), since disk dominated late-type galaxies predominate in the field, whereas bulge-dominate early-type galaxies are most common in clusters, one would have expected S0 galaxies in the field to, on average, be flatter than those in clusters. However, contrary to expectations, the data in Table 2 shows that this is not observed to be the case. This suggests that internal factors may have been important in the transformation of some spiral galaxies into lenticulars.

\section{CONCLUSIONS}

      The distribution of flattening values of S0 galaxies is found to be independent of environment. This result is surprising because one might have imagined the progenitors of S0 galaxies in clusters to have mainly been early-type (Sa) spirals, whereas it would have been expected that the majority of the ancestors of S0 galaxies in the field were late-type (Sc) spirals. Secondly, since massive luminous galaxies have deeper potential wells than low mass faint ones, one would have expected the luminosity distribution of lenticular galaxies to depend on environment. However, Table 1 suggests that this is not the case. No statistically significant differences are observed between the luminosity distributions of lenticular galaxies in clusters, groups and in the field. Thirdly, most present ideas on the formation of S0 galaxies do not explain why there are any lenticular galaxies in the field at all. This suggests that internal processes, such as the environmental effects produced by active galactic nuclei (Silk \& Norman 2009), might have been responsible for the absence of gas in some disk galaxies. Among northern Shapley-Ames galaxies 8\% are lenticulars that are located in the field. This number is significantly smaller than the 15\% of S0 galaxies in all clusters, and the 28\% S0 galaxies in the rich Virgo cluster. Finally, it is unexpected that the relative frequency of dusty $S0_{3}$ galaxies and of dust free $S0_1$ galaxies appears to be similar in clusters and in the field. Naively one might have expected that ram-pressure would have preferentially removed gas and dust from those lenticular galaxies that are located in clusters.
   
 The distribution of flattening values of elliptical galaxies in the Shapley-Ames sample does not differ significantly between clusters, groups and field ellipticals. In all three of these environments S0 galaxies are found to be $\sim$0.9 mag fainter than either E or Sa galaxies. This difference in luminosity might have been produced by the early sweeping of gas from the progenitors of S0 galaxies. Alternatively, it may have been easier to transform the low-luminosity spiral ancestors of present day S0 galaxies, than to transform the progenitors of more massive spirals into lenticular galaxies.
   
 In view of the many problems discussed above, the time may perhaps arrived, to rethink the nature of lenticular galaxies. Baade
(1963,p.77) wrote: ``Hubble had introduced the S0 type, in a way that was not easy to understand...I had great difficulty in understanding the matter, because he was trying to fit in S0 as a transition type between E galaxies and spirals, although he did not find it easy to show why there should be these transitional types.'' Perhaps we may we have to return to the view of Baade (1963,p. 78) who wrote: ``So, we should define it as the class of galaxies in which from their general form we should expect to find spiral structure, but which, contrary to our expectation, do not show it''. The main thrust of the present paper has been to show that a significant fraction of all S0 galaxies occur in the field where the effects of collisions and ram-pressure stripping are expected to have been minimal. This suggest that an internal, rather than an external, cause might have to be found for the absence of spiral structure in some field S0 galaxies. One might speculate that a weak (or absent) magnetic field (Chandrasekhar \& Fermi 1953ab) in proto-S0 galaxies was responsible for the lack of spiral arms in lenticular field galaxies. Alternatively, the present results might be interpreted as supporting the view of Larson et al. (1980) and of Bekki et al. (2002) that some S0 galaxies are objects that have been starved of inflowing external gas. In this respect they would differ from normal spirals that might still be fed by an external supply of gas from captured debris, minor mergers with gas-rich satellites, or even cold gas from cosmic filaments. A possible complicating factor is that the morphology of a galaxy might have been affected by both (1) its location in a cluster, and (2) by the presence of a nearby companion (Holmberg 1958, Hwang \& Park 2009). Finally, and perhaps most plausibly,
  it might be necessary to invoke internal processes (Silk \& Norman
2009) to account for the absence of significant amounts of gas in some field lenticular galaxies. As these authors point out, outflows from active galactic nuclei can trigger star formation via the compression of dense clouds. These star bursts would in turn produce supernovae that will enhance the outflow of gas from active nuclei.
   
 I thank Roberto Abraham for numerous useful discussions and Preethi Nair for making available her PhD thesis. Thanks are also due to Marie Martig for useful correspondence. I am also indebted to Brenda Parrish for technical support. Finally it is a pleasure to thank the referee for a very kind and helpful report.

\begin{deluxetable}{lrrrrrrrrr}
\tablecolumns{10}
\tablewidth{0pt}
\tablecaption{Luminosity distributions of cluster, group and field galaxies}
\tablehead{\colhead{} & \multicolumn{3}{c}{Ellipticals} & \multicolumn{3}{c}{Lenticulars}   &   \multicolumn{3}{c}{Sa}\\
\colhead{$M_{B}$}  & \colhead{C}  & \colhead{G}  & \colhead{F}  & \colhead{C}  & \colhead{G}  & \colhead{F}  & \colhead{C}  & \colhead{G}  & \colhead{F}}

\startdata

-23.25     &  1  &    0    &    0    &     0   &  0    &  0     &    0    &    0   &  0\\
-22.75     &  5  &    4    &    0    &     0   &  0    &  1     &    0    &    0   &  1\\
-22.25     &  6  &    1    &    1    &     0   &  1    &  1     &    2    &    1   &  3\\
-21.75     & 11  &    9    &    4    &     4   &  1    &  1     &    0    &    0   &  4\\
-21.25     & 12  &    2    &    8    &     5   &  3    &  4     &    8    &    2   &  7\\
-20.75     &  4  &    3    &    4    &     4   &  4    &  4     &    4    &    1   &  6\\
-20.25     &  3  &    1    &    0    &     7   &  4    &  3     &    6    &    0   &  4\\
-19.75     &  7  &    0    &    1    &    11   &  1    &  7     &    3    &    2   &  1\\
-19.25     &  3  &    0    &    1    &     7   &  2    &  2     &    1    &    1   &  0\\
-18.75     &  3  &    0    &    1    &     1   &  0    &  1     &    0    &    0   &  0\\
-18.25     &  1  &    0    &    0    &     1   &  0    &  0     &    0    &    0   &  0\\
-18.0      &  1  &    1    &    0    &     0   &  1    &  0     &    0    &    0   &  0\\
Total      &  57 &    21   &   20    &    40   &  17   &  24    &   24    &    7   &  26\\
Median     & -21.2 & -21.6 & -21.3   &   -20.1 & -20.5 & -20.3  &  -20.8  &  -20.5 & -21.1\\

\enddata

\end{deluxetable}

\begin{deluxetable}{rrrrr}
\tablecolumns{5}
\tablewidth{0pt}
\tablecaption{Flattening distribution of S0 galaxies in Virgo, in all clusters (including Virgo), in groups and in the field}
\tablehead{\colhead{e} & \colhead{Virgo} &\colhead{Clusters} &\colhead{Groups} & \colhead{Field}}
\startdata

0  &     1    &     4     &    2    &    3\\
1   &    0    &     0     &    0    &    2\\
2   &    2    &     2     &    1    &    0\\
3   &    3    &     6     &    2    &    1\\
4   &    1    &     1     &    1    &    2\\
5   &    0    &     2     &    3    &    4\\
6   &    2    &     4     &    3    &    2\\
7   &    3    &     4     &    0    &    4\\
8   &    2    &     4     &    3    &    5\\
9   &    1    &     6     &    1    &    1\\
10  &    1    &     2     &    0    &    0\\

\enddata

\end{deluxetable}

\begin{deluxetable}{lrrr}
\tablecolumns{4}
\tablewidth{0pt}
\tablecaption{Distribution of S0 subtypes}
\tablehead{\colhead{Environment} & \colhead{$S0_{1}$} &\colhead{$S0_{2}$} & \colhead{$S0_{3}$}}

\startdata

Virgo     &     11    &    0   &    4\\
Cluster   &     25    &    1   &    8\\
Group     &      9    &    2   &    4\\
Field     &     13    &    6   &    3\\

\enddata

\end{deluxetable}

\begin{deluxetable}{lrrrr}
\tablecolumns{5}
\tablewidth{0pt}
\tablecaption{Frequency distribution of morphological types in nothern Shapley-Ames galaxies}
\tablehead{\colhead{Type} & \colhead{Field} & \colhead{Groups} & \colhead{Clusters} & \colhead{Virgo}}

\startdata

E          & 20    &   21   &   61    &   11\\
E/S0       &  2    &    3   &   10    &    4\\
S0+SB0     & 33    &   19   &   59.5  &   24\\
S0/a+SB0/a &  7    &    3   &   12    &    3\\
Sa+SBa     & 37    &   11.5 &   33.5  &    5\\
Sab+SBab   & 17    &    2   &   19    &    7\\
Sb+SBb     & 58    &   23   &   36    &    7\\
Sbc+SBbc   & 51    &    8   &   32    &    2\\
Sc+SBc     &143    &   41.5 &   96.5  &   18\\
Scd+SBcd   &  4    &    2   &    7    &    1\\
Sd+SBd     &  0    &    2   &    9    &    1\\
Sm+SBm     &  5    &    1.5 &    5.5  &    1\\
Im+IBm     &  0    &    2   &    1    &    0\\
Other      & 14    &    5.5 &   12    &    1\\
Total     & 391    &  145   &  394    &   85\\

\enddata

\end{deluxetable}

\end{document}